# Phase field and level set methods for modeling solute precipitation and/or dissolution

Zhijie Xu[1,a)], Hai Huang[2], Xiaoyi Li[5], and Paul Meakin[2,3,4]

1. Idaho National Laboratory, Idaho Falls, Idaho 83415, USA.
   Now at Fundamental & Computational Science Directorate, Pacific Northwest National Laboratory, Richland, Washington 99352, USA.

2. Idaho National Laboratory, Idaho Falls, Idaho 83415, USA.

3. Physics of Geological Processes, University of Oslo, Oslo 0316, Norway.

4. Multiphase Flow Assurance Innovation Center, Institute for Energy Technology, Kjeller 2027, Norway.

5. United Technologies Research Center, United Technologies Research Center, East Hartford, Connecticut 06108, USA

The dynamics of solid-liquid interfaces controlled by solute precipitation and/or dissolution due to the chemical reaction at the interface were computed in two dimensions using a phase field models. Sharp-interface asymptotic analysis demonstrated that the phase field solutions should converge to the proper sharp-interface precipitation/dissolution limit. For the purpose of comparison, the numerical solution of the sharp-interface model for solute precipitation/dissolution was directly solved using a level set method. In general, the phase field results are found in good agreement with the level set results for all reaction rates and geometry configurations investigated. Present study supports the applications of both methods to more complicated and realistic reactive systems, including the nuclear waste release and mineral precipitation and dissolution.

Keyword: phase field, level set, precipitation, dissolution, reaction, diffusion.

a) Electronic mail: zhijie.xu@pnnl.gov

# 1. Introduction

Solute precipitation and/or dissolution at solid-liquid interfaces has a number of important practical applications including corrosion, etching, the formation of mineral deposits in boilers and heat exchangers, and in oil and water pipes, and the formation of gas hydrates in oil and gas pipelines. It also plays an important role in the rheology of the Earth's crust and subsurface science. For example, the growth of mineral precipitates in pore volumes and fracture apertures can cause a significant reduction in macroscopic porosity and permeability by plugging pore throats or increase porosity and permeability as a result of fracturing if the force of crystallization exceeds the rock strength. Similarly, reaction induced dissolution, enhanced by advection and diffusion, changes subsurface pore/grain geometries leading to concomitant changes in the macroscopic porosity and permeability. This is also an interesting topic in the study of nuclear waste released from the immobilized glass form. In the far field, the environment may be acidic due to the carbon dioxide dissolved in the seawater, or basic due to the surrounding clay or concrete. Dissolution rate of glass and the internal diffusion rate of radionuclides are significantly influenced by these environmental factors. Both rates are important in predicting the release rate of radionuclides from the glass form.

Mathematically similar to the material oxidation [1], solute dissolution and/or precipitation can be formulated as moving boundary problems. Historically these "Stephan problems" or moving boundary problems have been challenging from a computational point of view, but more recently a number of new computational methods have been developed to numerically solve them. The front-tracking [2], volume-of-fluid [3] and level set [4] methods are based on the tracking or capturing of sharp interfaces. The front-tracking algorithm can be used in conjunction with adaptive mesh refinement near the interface, but it is difficult to apply it to dynamic interfaces that undergo complicated topological changes. Volume-of-fluid (VOF) methods have the advantages of conserving mass by explicitly tracking volume fractions for each cell, but reconstruction of interfaces from volume fractions and calculation of associated geometric quantities, such as the interface curvature are not straightforward. This is particularly true in the 3-dimensional (3D) simulations. In general, the level set method can easily handle complex geometries, but it suffers from mass loss/gain problems if more complicated methods based on the volume-of-fluid or front tracking approaches [5], Lagrangian particles [6] or adaptive mesh refinement [7] are not implemented.

In contrast, the diffuse-interface models, such as the phase-field approach, do not require explicit computing of the moving interface. The approach, originally developed by van der Waals [8] in the 1800's and by Cahn and Hilliard [9] in the 1950's, is based on the concept of a diffuse interface that can be defined in terms of a density, structure or composition field (i.e. a phase field) that changes smoothly from one phase to the other over an interface zone with a non-zero width, $w$. In this manner, numerical difficulties associated with the boundary conditions at the sharp-interfaces are avoided, and no explicit interface tracking/capturing is required. The phase field is transported locally with the velocity of the interface and deformation of the phase field is restored by diffusive relaxation (Cahn-Hilliard [9]) of conservative phase fields or direct relaxation (Allen-Cahn [10]) of non-conservative fields. Originally proposed for the applications such as the solidification of pure melts[11-15], the phase field approach has been used to simulate a wide variety of interface dynamics phenomena including two-phase Navier-Stokes flow [16], crack propagation and dislocation dynamics [17], and chemical reaction induced precipitation and/or dissolution [18, 19].

However, in most applications, the phase field equations are only used to circumvent the difficulty of tracking sharp interfaces, where the equations and/or their associated parameters do not directly represent the physics of the interfaces. Phase field methods have also been used in conjunction with particle models such as smoothed particle hydrodynamics (SPH) and dissipative particle dynamics (DPD), either to calculate forces based on a free energy functional [20], or more precisely locate complex boundaries [21] as demonstrated in our review article [22]. This type of hybrid method of phase field and particle models has also been applied in the study of biofilm growth kinetics [23].

In this paper, we first describe the level set and the phase field models used for the physics of solute precipitation/dissolution at the solid-liquid interface. We then present quantitative numerical simulations of precipitation/dissolution, using both phase field and level methods for the same problems, and cross-validate both approaches by comparing their predictions.

## 2. Sharp-interface model and phase-field equations

### 2.1. Sharp-interface model and level set method

The sharp interface model provides the governing equations for the solute precipitation/dissolution free-boundary problem. The simplest model includes diffusion in the liquid and first order reactions at the liquid-solid interface. The dynamics of the moving interface during precipitation/dissolution is, in general, controlled by the rate of the transport of dissolved solute away from or to the interface (diffusion in the solid phase is usually small enough to be neglected) and the reaction rate at the interface. The system of equations for this diffusion-precipitation/dissolution problem is given by

$$\partial c / \partial t = D \nabla^2 c, \tag{1}$$

$$c|^+ = v_s / bkk_c \text{ on } \Gamma, \tag{2}$$

$$v_s = bk_c D \nabla c|^+ \cdot \mathbf{n} \text{ on } \Gamma, \tag{3}$$

where $D$ is the solute diffusion coefficient. $c = (C - C_e^\infty)/C_e^\infty$ is the normalized solute concentration, where $C(\mathbf{x}, t)$ is the solute concentration at position $\mathbf{x}$ and time $t$ and $C_e^\infty$ is the solute concentration at equilibrium with the solid at a planar interface. Here, $\Gamma$ represents the solid-liquid interface. $c|^\pm$ and $\nabla c|^\pm$ are the dimensionless solute concentrations and concentration gradients at the interface with $|^+$ indicating the magnitude of a variable at the liquid side of the interface and $|^-$ indicating the magnitude at the solid side in the sharp-interface model. The dimensionless solute concentration on the solid side satisfies $c|^- = 0$ and $\nabla c|^- = \mathbf{0}$). The dimensionless variable $b$ is defined as $b = C_e^\infty / \rho_s$ ($\rho_s$ is the density of the solid). Equation (2) describes the chemical reaction kinetics at the solid-liquid interface, and Eq. (3) expresses local mass conservation. $v_s$ is the velocity of the interface in the direction normal to the interface, and $k_c$ is a stoichiometric coefficient of order unity [24]. In Eq. (3), $\mathbf{n}$ is the unit vector perpendicular to the interface, $\Gamma$, pointing into the liquid and $k$ is the reaction rate coefficient.

Equations (1)-(3) can be rewritten in dimensionless form by introducing three dimensionless constants, $\alpha = 1/2bk_c$, Péclet number $P_e = UL/D$ and Damköhler number $D_a = kL/D$, where $L$ is a characteristic length, $U$ is a characteristic velocity, and $L/U$ serves as a characteristic time. The resulting dimensionless equations are:

$$\partial c / \partial t = \nabla^2 c / P_e, \tag{4}$$

$$c\big|^+ = (2\alpha P_e / D_a) v_s \text{ on } \Gamma, \tag{5}$$

$$v_s = (1/2\alpha P_e) \nabla c\big|^+ \cdot \mathbf{n} \text{ on } \Gamma. \tag{6}$$

Li et al. [25] applied a level set interface capturing approach to model this coupled moving boundary problem. Level set interface capturing is based on the idea that the solid-fluid interface, $\Gamma$, in $d$-dimensional space can be represented by a cut through a surface in ($d$+1)-dimensional space (like a two-dimensional shoreline is a cut through the three-dimensional land surface at sea level). In the level set method, the $d$-dimensional interface, $\Gamma(\mathbf{x}) = \Gamma(x_1, x_2, \ldots x_d)$, is defined as the zero level contour (zero contour line in 2D or zero isosurface in 3D) or cut through an evolving ($d$+1)-dimensional level set field $\varphi(\mathbf{x}, t)$ as $\Gamma = \{\mathbf{x} \mid \varphi(\mathbf{x}, t) = 0\}$. In principal, a wide range of level set fields can be used, providing that $\varphi > 0$ for phase 1, $\varphi < 0$ for phase 2, and $\varphi = 0$ at the interface. In practice the signed distance function $\varphi(\mathbf{x}, t)$ was used, which has an absolute value equal to the minimum distance from the point *x* to the interface, and a sign that is positive in phase 1 and negative in phase 2. $\varphi$ has a gradient of unity near the interface and is used to enable more accurate calculation of geometric quantities such as interface norms and curvatures. The unit vector normal to the interface, *n*, can be obtained by simply determining the gradient of $\varphi$ at the interface. Due to chemical reactions, the interface evolves and the evolution of the signed distance function $\varphi$ is governed by an advection equation as $\dfrac{\partial \varphi}{\partial t} + u_n^\Gamma \mathbf{n} \cdot \nabla \varphi = 0$, where $u_n^\Gamma$ is the interface velocity along the normal direction.

System of equations (Eqs. (4)-(6)) can be solved using a standard level set interface capturing algorithm by repeating following steps:

(a) advancing the level set function using solid-liquid interface velocities (due to precipitation/ dissolution) extrapolated from the velocities calculated from Eq. (6);

(b) reinitialize the level set function to a signed distance function;

(c) calculate the new concentration field;

In practice, the level set function is computed only in a narrow band that contains the interface region to reduce the computational burden. Detailed presentations of level set methods can be found in the literature [4] and the numerical algorithms used in this work are described in Li et al. elsewhere [25, 26]. We choose $P_e = 1$ and $\alpha = 0.5$ for all level set simulations. The other variables are the grid resolution, $N$, and the Damköhler number, $D_a$ and the corresponding boundary conditions for the concentration field.

## 2.2. Phase field method

The phase field equations for the dynamics of liquid-solid interfaces that evolve due to precipitation/dissolution together with a rigorous asymptotic analysis to show that the phase field solutions converge to the proper sharp-interface limit have already been presented [18]. The phase field equations are,

$$\tau \frac{\partial \phi}{\partial t} = \varepsilon^2 \nabla^2 \phi + (1-\phi^2)(\phi - \lambda c) - \varepsilon^2 \kappa |\nabla \phi| \quad (7)$$

$$\frac{\partial c}{\partial t} = D\nabla^2 c + \alpha \frac{\partial \phi}{\partial t}\left(1 + \frac{D\nabla^2 \phi - \partial \phi/\partial t}{k|\nabla \phi|}\right), \quad (8)$$

$$\tau = \alpha\lambda \frac{\varepsilon^2}{D}\left(\frac{5}{3} + \frac{\sqrt{2}D}{k\varepsilon}\right), \quad (9)$$

where $\phi(\mathbf{x},t)$ is the phase-field variable at position *x* and time *t*. The phase field variable, $\phi$, varies from -1 to 1 with $\phi \approx -1$ indicating that *x* lies in the solid phase, $\phi \approx +1$ indicating that *x* lies in the liquid phase and other values *x* (-1+Δ<$\phi$<1-Δ, where Δ is small but not infinitesimally small) indicating *x* lies in the diffuse interface zone. The quantities $\tau$, $\varepsilon$ and $\lambda$ introduced in Eqs. (7)-(9) are all microscopic phase field parameters ($\tau$ is a characteristic time parameter, $\varepsilon$ is a parameter that is closely related to the interface thickness and $\lambda$ is a dimensionless parameter that controls the coupling strength between the phase field variable $\phi$ and the concentration field *c*). $\kappa = \nabla \cdot n$ represents the interface curvature and $n = \nabla\phi/|\nabla\phi|$ is the interface normal. *D* and *k* are the diffusion and reaction rate constants, the same as those in the sharp interface model (Eqs. (1)-(3)). The relationship between $\tau$, $\varepsilon$, and $\lambda$ (Eq. (9)) is obtained from the formal asymptotic analysis [18].

Similarly, the phase field equations are expressed in dimensionless form by introducing the units of length *L*, time $L/U$, and two dimensionless constants - the Péclet number $P_e = UL/D$ and the Damköhler number $D_a = kL/D$. The phase field equations can then be rewritten as (the prime is dropped for simplicity),

$$\frac{\partial \phi}{\partial t} = \frac{1}{P_e^\phi}\left\{\nabla^2\phi + \chi^2\left(1-\phi^2\right)(\phi-\lambda c) - \kappa'|\nabla\phi|\right\}, \tag{10}$$

$$\frac{\partial c}{\partial t} = \frac{1}{P_e}\nabla^2 c + \alpha\frac{\partial \phi}{\partial t} + \left(\nabla^2\phi - P_e\,\partial\phi/\partial t\right)\frac{\alpha\cdot\partial\phi/\partial t}{D_a|\nabla\phi|}, \tag{11}$$

$$P_e^\phi = P_e\,\tau D/\varepsilon^2, \quad \kappa' = \kappa L, \quad \chi = L/\varepsilon, \tag{12}$$

$$\lambda = P_e^\phi\left/\left[P_e\alpha\left(5/3+\sqrt{2}\chi/D_a\right)\right]\right.. \tag{13}$$

These equations can be further simplified to

$$\frac{\partial \phi}{\partial t} = \frac{1}{P_e^{'}}\left\{\nabla^2\phi + \left(1-\phi^2\right)\left(\phi - \lambda c\right) - \kappa^{'}\left|\nabla\phi\right|\right\}, \tag{14}$$

$$\frac{\partial c}{\partial t} = \nabla^2 c + \alpha\frac{\partial \phi}{\partial t} + \left(\nabla^2\phi - \partial\phi/\partial t\right)\frac{\alpha \cdot \partial\phi/\partial t}{D_a\left|\nabla\phi\right|}, \tag{15}$$

$$P_e^{'} = \tau D/\varepsilon^2 \,,\ \kappa^{'} = \varepsilon\kappa, \tag{16}$$

$$\lambda = P_e^{'}\Big/\left[\alpha\left(5/3 + \sqrt{2}/D_a\right)\right]. \tag{17}$$

by simply taking the unit of length $L = \varepsilon$ and unit of time $L/U = \varepsilon^2/D$, (or equivalently setting $P_e = \chi = 1$ by choosing $L = \varepsilon$ and $U = D/\varepsilon$)

The original moving boundary problem is reduced to coupled partial differential Eqs. (14) and (15) with three dimensionless parameters $\alpha$, $P_e^{'}$, and $D_a$. In this manner, the computational difficulties associated with front tracking are avoided because the interface is not explicitly tracked in the phase field formulation. The parameter $P_e^{'}$ is relevant to the diffusion of the phase field variable $\phi$. A small $P_e^{'}$ means fast diffusion of $\phi$. In all phase field simulations, we set $P_e^{'} = 1$ (i.e. $\tau = \varepsilon^2/D$) without losing generality and a single time step $\Delta t$ can be used to advance both $\phi$ and $c$ fields using an explicit scheme. Equations (14) and (15) were solved numerically on a two-dimensional rectangular lattice with a constant grid spacing, $\Delta x$, in both directions. Sufficient accuracy with a reasonable computational load was achieved using a grid spacing in the range $0.25\varepsilon \leq \Delta x \leq 0.5\varepsilon$ [27].

We choose $\Delta x = 0.5\varepsilon$ and $\alpha = 0.5$ for all phase field simulations. The other variables are the phase field parameter, $\varepsilon$, (or equivalently the grid resolution, $N = 1.0/\varepsilon$), and the Damköhler number, $D_a$, the same set of parameters required for the level set simulations. Therefore, a direct comparison can be made between the two methods.

The Laplacian in both the phase field and level set simulations is computed from the commonly used 5-point finite-difference stencil,

$$\nabla^2 \phi_{i,j} = \frac{\phi_{i+1,j} + \phi_{i-1,j} + \phi_{i,j+1} + \phi_{i,j-1} - 4\phi_{i,j}}{\Delta x^2}, \tag{18}$$

and the normal of gradient of $\phi$ is computed using the central difference approximation,

$$|\nabla \phi|_{i,j} = \frac{1}{\Delta x}\sqrt{\frac{(\phi_{i+1,j} - \phi_{i-1,j})^2}{4} + \frac{(\phi_{i,j+1} - \phi_{i,j-1})^2}{4}}. \tag{19}$$

The curvature term in Eq. (14) is computed from the curvature equation $\kappa' = \nabla \cdot (\nabla \phi / |\nabla \phi|)$. Time discretization is implemented using the first order forward Euler method. The time step used for the time integration satisfied the constraints from the numerical stability required by the explicit integration of both phase field $\phi$ and concentration field $c$ (i.e. the Courant–Friedrichs–Lewy condition). The two-dimensional numerical schemes presented here can be easily extended to three-dimensional phase field simulations.

## 3. Results and discussion

We first compute the symmetric circular growth in a simulation domain consisting of 0.5 × 0.5 square box (one quadrant of a 1.0 × 1.0 box), which can be replicated to fill the 1.0 × 1.0 box using the 4-fold rotational symmetry of the lattice, using both the level set and the phase field methods with the same resolution. As shown in Fig. 1, solute precipitation and/or dissolution were initiated with a quarter disk of radius $r_0$ placed in the lower-left corner of the computational domain. Various Damköhler numbers, $D_a = 0.1$, 1.0 and 10.0, are used to represent slow to fast chemical reaction rates compared to the solute diffusion. The far field concentration, $c_\infty$, was prescribed on the circular boundary of radius 0.5. To ensure the grid convergence, simulations were performed with $r_0 = 0.1$, $D_a = 1.0$ and $c_\infty = 1.0$ at various grid resolutions ($N$ = 50, 100 and 200, or equivalently $\Delta x$ = 0.01, 0.005 and 0.0025). The instantaneous solid area $A$ was computed during both level set and the phase field simulations and Fig. 2 shows the area, $A$, at four different times (t = 0.1, 0.2, 0.3 and 0.4) for different grid resolutions and simulation methods. The numerical results from the phase field and level set methods are in a very good agreement with each other and the error is reduced as finer grids are used (they appear to converge to the same values as $\Delta x \rightarrow 0$).

An initial radius of $r_0 = 0.1$ was used for the solute precipitation simulations. Figure 3 shows a comparison between phase field and level set methods for various Damköhler number, $D_a$ = 0.1, 1.0 and 10.0, with a far-field solute concentration of $c_\infty = 0.1$, and Fig. 4 shows a comparison between the phase field and level set methods with a far-field solute concentration $c_\infty = 1.0$. For both boundary conditions, the phase field simulation results are in good agreement with the level set method results. The rate of solid precipitation increases significantly with increasing Damköhler number, $D_a$, and with increasing far field concentration $c_\infty$.

In order to investigate dissolution, solute dissolution simulations were performed with an initial radius of $r_0 = 0.4$ and far field concentrations of $c_\infty$ = -0.1 and -1.0 (a negative $c_\infty$ means that the absolute solute concentration is less than the equilibrium concentration, and this leads dissolution of the solid). Figures 5 and 6 compare the reduction of the solid area $A$ obtained from the phase field and level set models for Damköhler number of $D_a$ = 0.1, 1.0 and 10.0 and two different far field concentrations ($c_\infty = -0.1$ and $c_\infty = -1.0$). Again, the phase field results are in good agreement with the level set method for all Damköhler numbers and far field concentrations.

To make the comparison more comprehensive and convincing, we also studied the growth of a non-circular shape with positive (convex) and negative (concave) curvatures. The equation describing the initial shape of the solid in polar coordinate system is,

$$r = 0.1 + 0.025\cos(4.0\theta), \tag{20}$$

where $r$ is the radial coordinate and $\theta$ is the angular coordinates. This perturbed shape intercepts the $x$ and $y$ axis at a distance of $r = 0.125$ and intercepts the line $y=x$ with a distance of $r = 0.075$ from the origin. The evolution of the solid-liquid interfaces is illustrated in Fig. 7. The phase field and level set results are represented by the solid lines and dashed lines, respectively. Again, a satisfactory agreement between phase field and level set simulation results was obtained.

## 4. Conclusion

The phase field and level set methods have been used to compute the time dependent solution of two-dimensional solute precipitation/dissolution problem with dynamically evolving solid-liquid interfaces. The phase-field approach does not require explicit computing of the moving interface so that numerical difficulties associated with the moving interface are avoided and no explicit interface tracking/capturing is required. However, phase field approach requires a complex asymptotic analysis in order to find the relation between parameters of the physical model and the phase field method [18]. Level set approach does not require such an asymptotic analysis and can naturally handle the discontinuities at the interface. An explicit interface tracking is required for this method.

Both methods were implemented for circular and perturbed shape at various Damköhler numbers and far-field concentrations. The good agreement between the numerical results obtained using these two methods cross-validated both of them and supports the practical application of both methods to more realistic systems. However, our implementation of both methods is not efficient and practical applications (for example, nuclear waste release and mineral precipitation and/or dissolution) will require the incorporation of advanced adaptive techniques. In order for an efficient implementation for practical applications, both methods require a much finer mesh or grid around the solid-liquid interface and a coarse mesh in other regions. For example, local adaptive mesh refinement can be used to efficiently resolve the thin interface and this adaptive refinement can be repeated dynamically to ensure that the thin interface region is covered with the finest mesh during the entire simulation time.

Acknowledgements:

This work was partially supported by the U.S. Department of Energy, Office of Science Scientific Discovery through Advanced Computing Program, and partially funded by the US Department of Energy's Nuclear Energy Advanced Modeling and Simulation (NEAMS) program. The Pacific Northwest National Laboratory is operated by Battelle Memorial Institute for the US Department of Energy under contract No. DE-AC05-76RL01830. The Idaho National Laboratory is operated for the U.S. Department of Energy by the Battelle Energy Alliance under Contract DE-AC07-05ID14517.

FIG. 1. Schematic plot of the simulation domain

FIG. 2. Comparison of areas at t = 0.1, 0.2, 0.3 and 0.4, obtained using the phase field (PF) and level set (LS) models for circular growth with various grid spacing $\Delta x = 0.01$, 0.005, and 0.0025, corresponding to grid resolutions of $N$ = 50, 100, and 200. The simulations were performed with a Damköhler number of $D_a = 1.0$ and a far field concentration of $c_\infty = 1.0$.

FIG. 3. Comparison of the growth of the solid area obtained from phase field (PF) and level set (LS) simulations of circular growth due to precipitation with Damköhler numbers of $D_a$ = 0.1, 1.0 and 10.0 and a far field concentration of $c_\infty = 0.1$.

FIG. 4. Comparison of the growth of the solid area obtained from phase field (PF) and level set (LS) simulations of circular growth due to precipitation with Damköhler numbers of $D_a$ = 0.1, 1.0 and 10.0 and a far field concentration of $c_\infty = 1.0$ (ten times larger than that used to obtain the results shown in Fig. 3).

FIG. 5. Comparison of the reduction of the solid area obtained from phase field (PF) and level set (LS) simulations of circular dissolution with Damköhler numbers of $D_a$ = 0.1, 1.0 and 10.0 and a far field concentration $c_\infty = -0.1$.

FIG. 6. Comparison of the reduction of the solid area obtained from phase field (PF) and level set (LS) for circular dissolution with Damköhler numbers of $D_a$ = 0.1, 1.0 and 10.0 and a far field concentration $c_\infty = -1.0$.

FIG. 7. Comparison of phase field (PF) and level set (LS) simulations for the growth of a not circular shape with concave and convex curvatures with a Damköhler number of $D_a$ = 1.0 and far field concentration of $c_\infty = 1.0$.

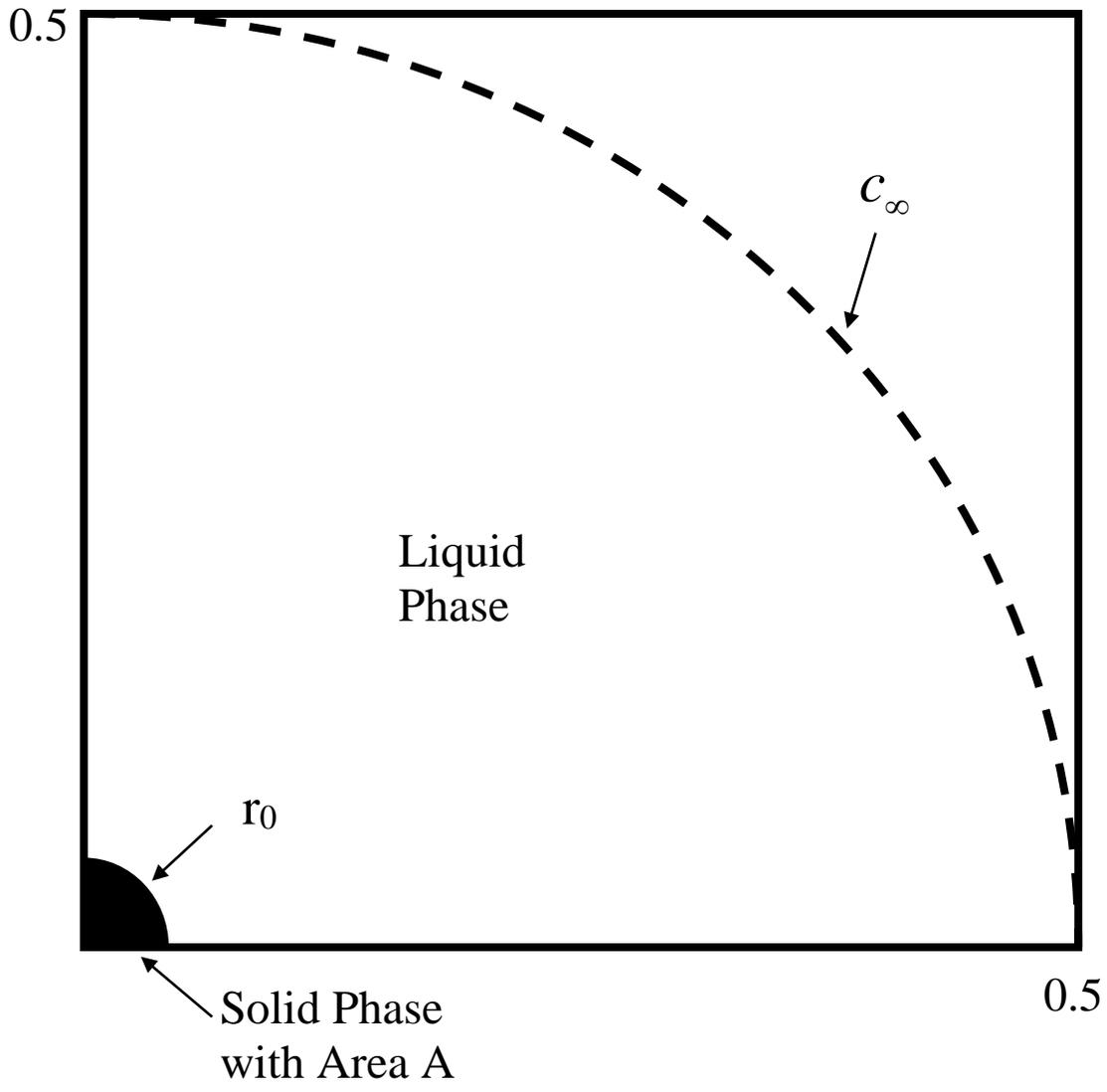

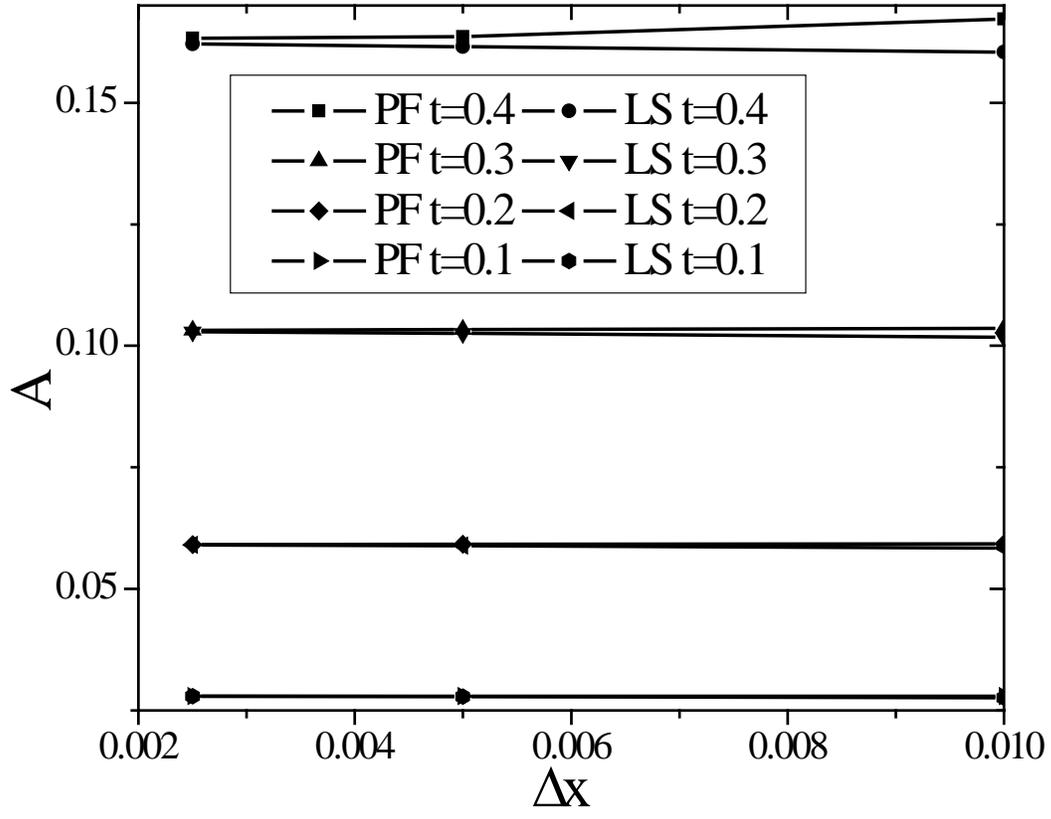

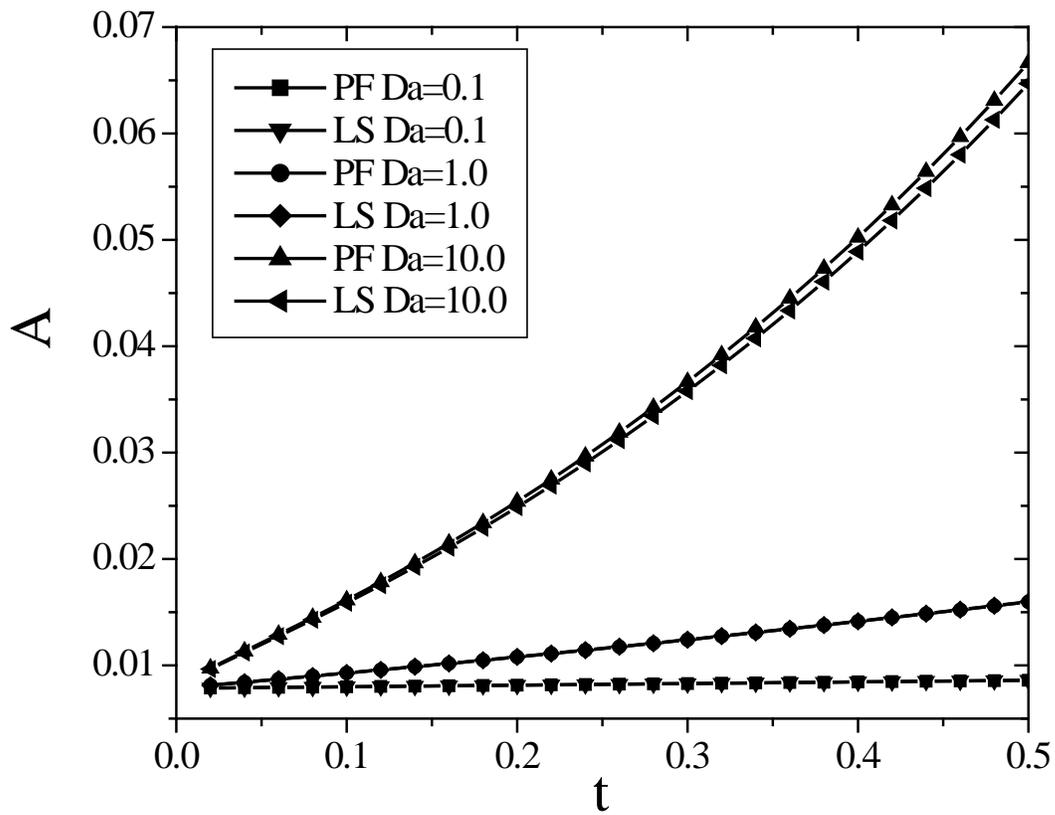

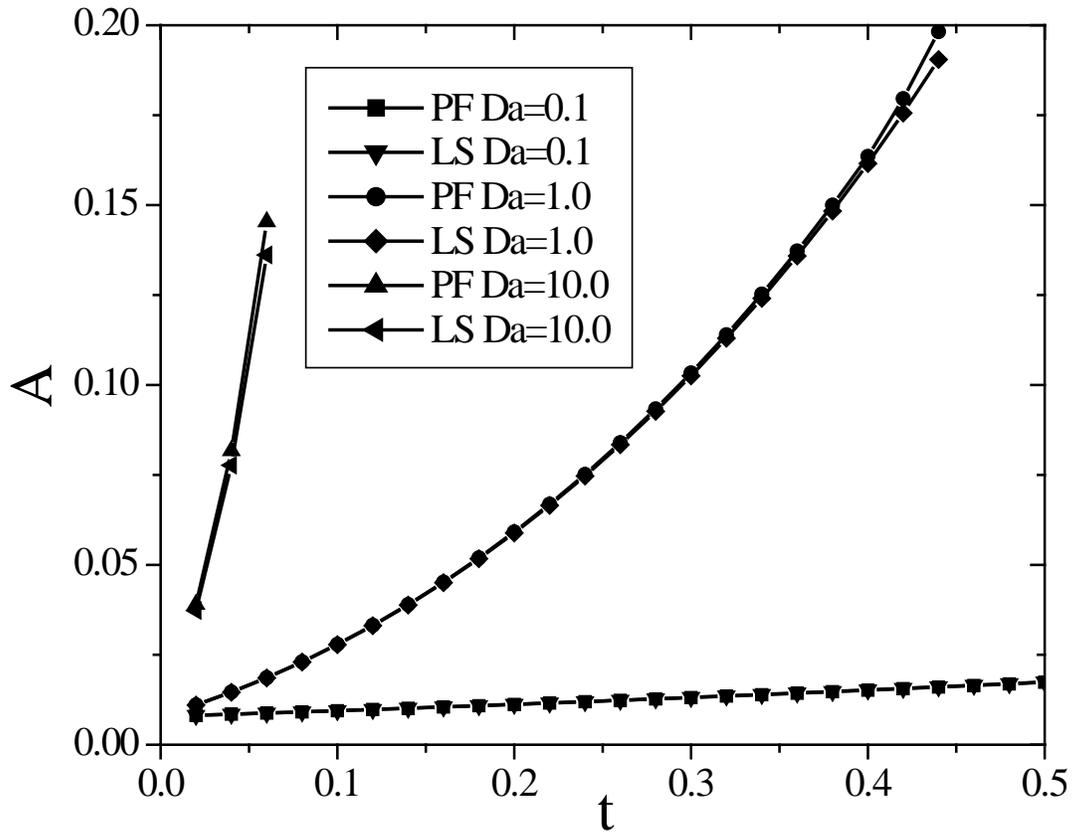

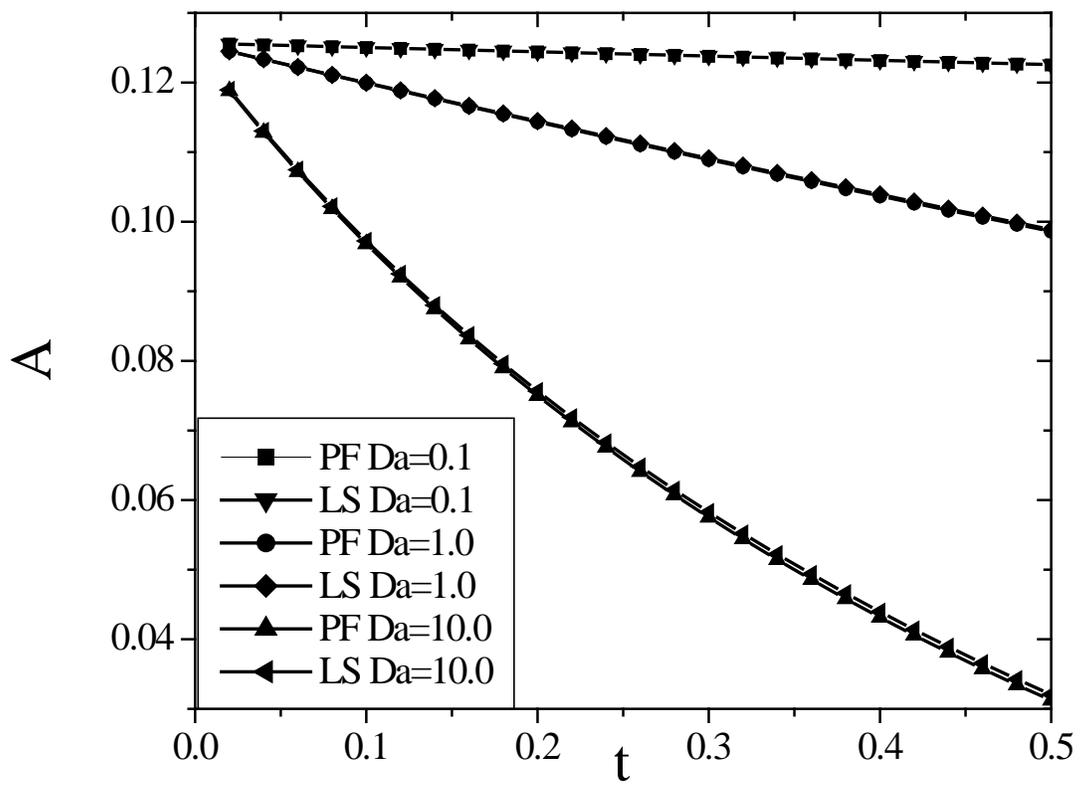

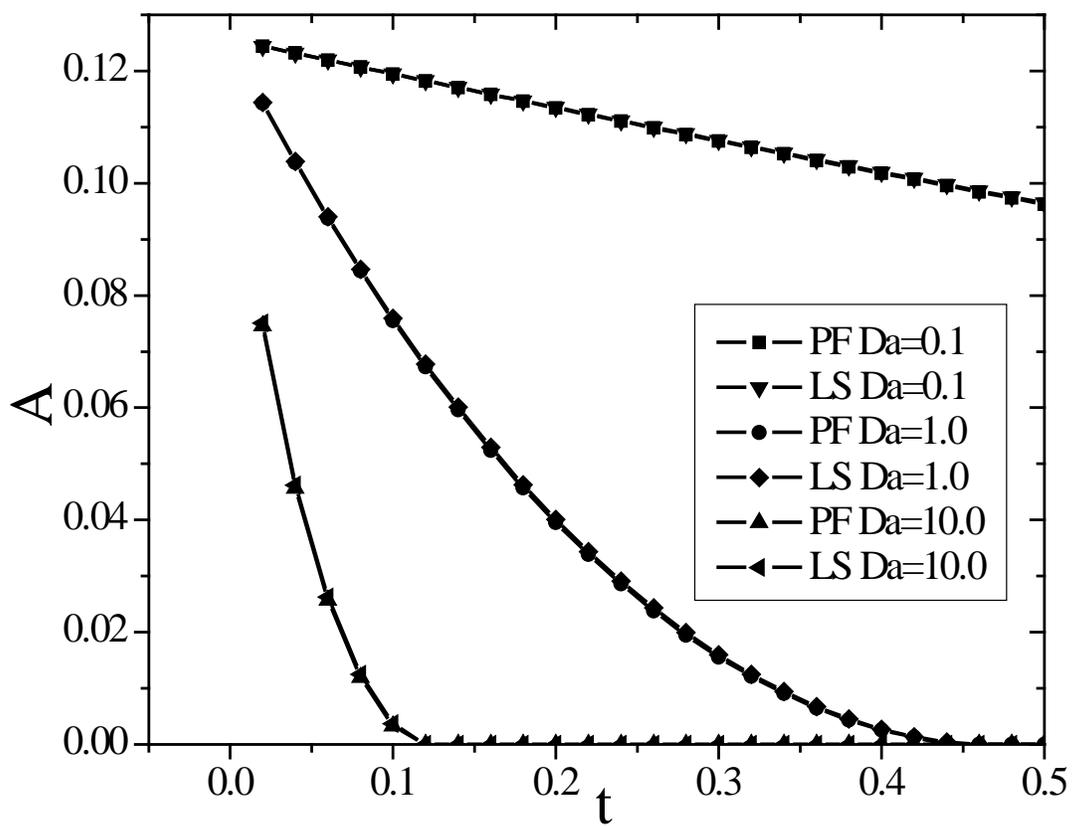

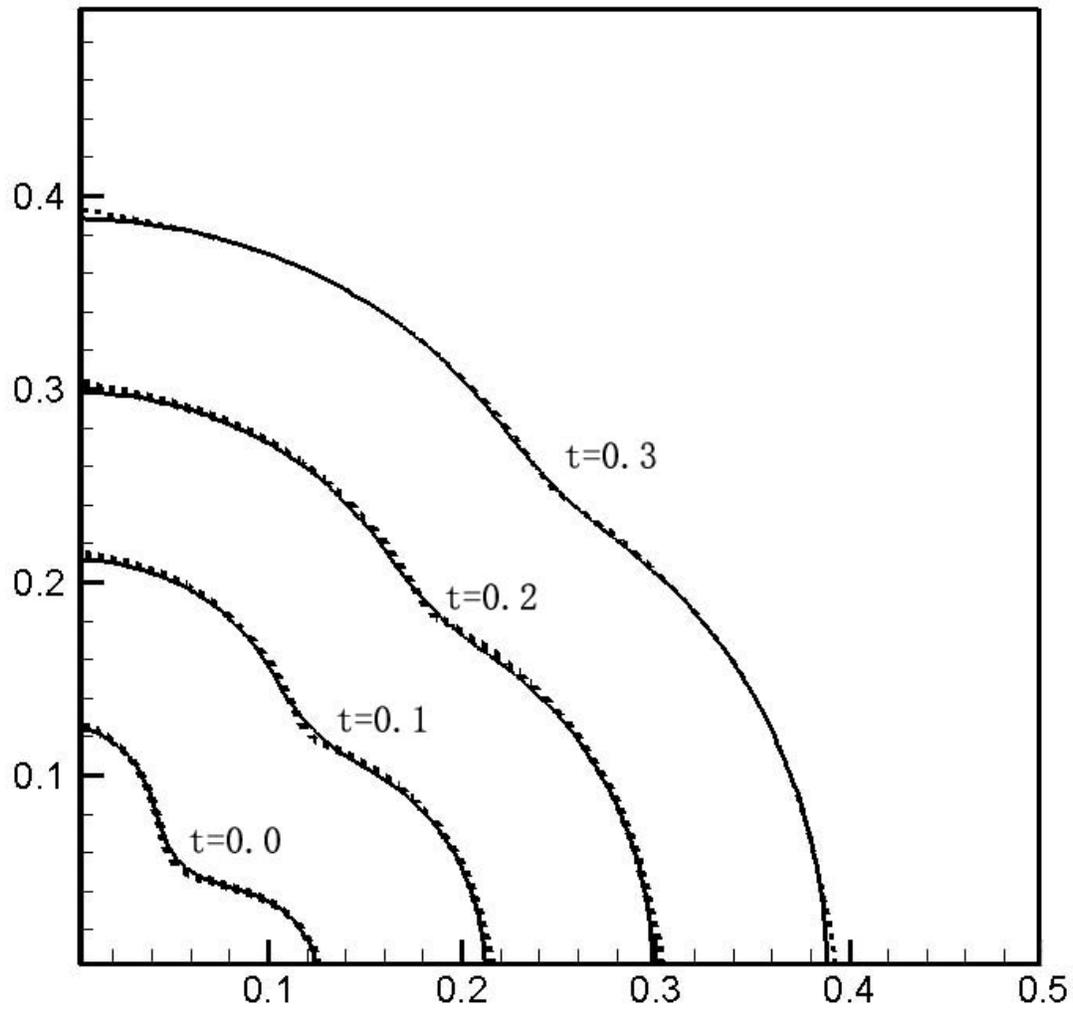

# References

[1] Z. Xu, K.M. Rosso, S.M. Bruemmer, J. Chem. Phys., 135 (2011) 024108.
[2] J. Glimm, J.W. Grove, X.L. Li, K.M. Shyue, Y.N. Zeng, Q. Zhang, Siam Journal on Scientific Computing, 19 (1998) 703-727.
[3] R. Scardovelli, S. Zaleski, Annu. Rev. Fluid Mech., 31 (1999) 567-603.
[4] S. Osher, J.A. Sethian, J. Comput. Phys., 79 (1988) 12-49.
[5] M. Sussman, K.M. Smith, M.Y. Hussaini, M. Ohta, R. Zhi-Wei, J. Comput. Phys., 221 (2007) 469-505.
[6] D. Enright, R. Fedkiw, J. Ferziger, I. Mitchell, J. Comput. Phys., 183 (2002) 83-116.
[7] R.R. Nourgaliev, S. Wiri, N.T. Dinh, T.G. Theofanous, Int. J. Multiph. Flow, 31 (2005) 1329-1336.
[8] J.D. van der Waals, Verhandel. Konink. Akad. Weten, (Amsterdam, 1893).
[9] J.W. Cahn, J.E. Hilliard, J. Chem. Phys., 28 (1958) 258-267.
[10] S.M. Allen, J.W. Cahn, Acta Metall., 27 (1979) 1085-1095.
[11] A. Karma, W.J. Rappel, Phys. Rev. E, 53 (1996) R3017-R3020.
[12] J.B. Collins, H. Levine, Phys. Rev. B, 31 (1985) 6119-6122.
[13] J.S. Langer, Directions in Condensed Matter, (World Scientific, Philadelphia, 1986).
[14] C. Beckermann, H.J. Diepers, I. Steinbach, A. Karma, X. Tong, J. Comput. Phys., 154 (1999) 468-496.
[15] D.M. Anderson, G.B. McFadden, A.A. Wheeler, Physica D, 135 (2000) 175-194.
[16] D. Jacqmin, J. Comput. Phys., 155 (1999) 96-127.
[17] L.Q. Chen, Annu. Rev. Mater. Res., 32 (2002) 113-140.
[18] Z. Xu, P. Meakin, J. Chem. Phys., 129 (2008) 014705.
[19] Z. Xu, P. Meakin, J. Chem. Phys., 134 (2011) 044137.
[20] Z. Xu, P. Meakin, A.M. Tartakovsky, Phys. Rev. E, 79 (2009) 036702.
[21] Z. Xu, P. Meakin, J. Chem. Phys., 130 (2009) 234103.
[22] P. Meakin, Z.J. Xu, Prog. Comput. Fluid Dyn., 9 (2009) 399-408.
[23] Z.J. Xu, P. Meakin, A. Tartakovsky, T.D. Scheibe, Phys. Rev. E, 83 (2011).
[24] B. Berkowitz, J.Y. Zhou, Water Resour. Res., 32 (1996) 901-913.
[25] X.Y. Li, H. Huang, P. Meakin, Water Resour. Res., 44 (2008) -.
[26] X.Y. Li, H. Huang, P. Meakin, Int. J. Heat Mass Transfer, 53 (2010) 2908-2923.
[27] Y. Sun, C. Beckermann, J. Comput. Phys., 220 (2007) 626-653.